# Design and implementation of brain surgery bipolar electrocautery simulator using haptic technology

Reza Karimzadeh[1,2], Javad Sheikh[3], Hamed Azarnoush[1], and Hossein Arabi[4]

[1]Department of Biomedical Engineering, Amirkabir University of Technology (Tehran Polytechnic), Tehran, Iran

[2]Department of Electrical Engineering, Sharif University of Technology, Tehran, Iran

[3]Department of Computer Engineering, Amirkabir University of Technology (Tehran Polytechnic), Tehran, Iran

[4]Division of Nuclear Medicine and Molecular Imaging, Geneva University Hospital, CH-1211 Geneva 4, Switzerland

[†]**Corresponding Author:**

Hamed Azarnoush, Ph.D.
Amirkabir University of Technology (Tehran Polytechnic)
Department of Biomedical Engineering
No. 350, Hafez Ave, Valiasr Square, Tehran, Iran 1591634311
**Tel:** +98-9128933423
**email:** hamed.azarnoush@gmail.com

**Abstract**

Surgical simulators have been widely used in training and evaluation of physicians and surgeons. Virtual reality augmented with haptic technology has made it feasible to develop more realistic surgical simulators. In this context, we set out to design and develop a brain surgery bipolar electrocautery simulator using haptic technology. A 3D model of brain tissue was generated based on a brain craniotomy image. Bipolar forceps were also modeled to visually assimilate real forceps. An experiment was developed to assess the learning process of the participants. In this experiment, the volunteers were asked to cauterize a large blood vessel in the brain while minimizing the damage done to the brain tissue. The experiment was performed on 20 volunteers, and statistical analysis was conducted on the learning process and error reduction during the surgery. Next, the volunteers were divided into gamer and non-gamer groups. The analysis of the volunteers' operation demonstrated that, on average, there was a 5 percent reduction in the percentage of applied force error. It was also shown that the results achieved by the gamer and non-gamer group has significant difference with a p-value of 0.0001. So, playing computer games would increase hand control, focus, and reflex and positively affect surgery skills.

# 1   Introduction

A three-year study by the US Health and Drug Administration found that medical errors caused death of more than 230,000 people in US hospitals [1]. Surgical errors cause various problems as 59.2% of surgical error victims suffer from temporary injuries, 32.9% suffer from permanent, irreparable injuries, and 6.6% lead to death [2]. It is not possible to determine the exact number of surgical errors because most hospitals restrain to disclose information regarding surgical errors. Over the past decades, in order to increase patient safety in the operating room there has been more focus on supportive practices such as devices, drugs, staff, and management of practices. However, there has been less focus on surgeon's techniques and operations in the operating room [3].

Today, the focus is starting to shift toward training and evaluating the surgeon in simulation environments. One approach to training is using plastic models, anesthetized animals, or human corpses to gain skills. However, plastic models are not usable for electrocautery. The bodies of animals and humans, besides the moral and legal barriers, are usually not readily available in training settings. Moreover, for certain reasons such as the differences between human and animal anatomy and blood pressure in living organisms, their use has limited effectiveness for electrocautery simulation [1, 4].

In recent decades, several studies have employed computer tools to educate and train residents, and these studies have received positive results for the training of the residents [5, 6, 7]. Virtual reality has also been used to train residents in laryngoscopy, photocoagulation, dermatologic, and endoscopy surgery [8, 9, 10, 11, 12]. In one experiment, two groups of apprenticeship trainees were examined. The group that used virtual laparoscopic cholecystectomy simulators in their program performed faster with less errors compared to those who did not use this simulator [13]. Another study with 612 participants showed that a virtual laparoscopy simulator decreased the surgery time, increased accuracy, and reduced error rate [14]. It has also been proposed that virtual simulation can be used to assess surgical skill proficiency [15, 16].

Several reports have also been published on Virtual Reality (VR) neurosurgery simulations. One of these simulators is Vivian, which does not allow force manipulation on brain matter. However, the ImmersiveTouch system simulates haptic feedback using a hand and head tracker system. Furthermore, the NeuroTouch system simulates brain surgeries, including tumor resection [15].

We developed a VR simulator in which an operator could practice using a bipolar electrocautery and coagulate brain vessels. Reports of various surgical device simulators have been published. However, no simulation results related to cerebral vascular coagulation tasks have been reported. Nowadays, bipolar electrocautery devices have become an essential tool for surgeons. In this device, a current passing from one pole to the other coagulates the blood flow path and prevents bleeding when tissue is cut by clotting blood vessels. In this project, bipolar coagulation forceps and their interaction with brain matter during cerebral vascular coagulation have been modeled to potentially teach and evaluate the duration and force needed during the surgery. The participants achieved better performance and speed after some training, and the evaluation metrics showed clear improvement too. This research utilizes a haptic device named Geomagic Touch (3D Systems, Rock Hill, South Carolina, USA), to simulate this surgical task. This device was made in a 3-dimension mode which simulates the interaction of the virtual environment in real-time. Through connection to the computer and receiving the information from the virtual environment, it transmits the simulated tactile force, related to interaction with virtual objects, to the user.

The contributions of this work are summarized as follows:

1- A surgical simulator for electrocautery is developed
2- Evaluation metrics for assessing the performance of the participants are created
3- The effect of gaming on surgical skill is studied

Section 2 describes the methods used to design and develop the simulator together with experiments to evaluate the learning process using this simulator. Section 3 includes the results which analyze the experiment data obtained from volunteers and their progress in vascular coagulation task.

## 2   Methods

In this section, the design of the simulator is described and how different parts of this simulator like the brain tissue, tools, and their reaction to touch and electrocautery are simulated. Next, the vein coagulation experiment and the evaluation metrics for this operation are described.

### 2.1   Simulator

In the first step, a virtual environment is created, this virtual space needs to be illuminated through creating a virtual light source, and a camera is required to simulate the user's eyesight. After creating these two elements, a haptic loop and a graphic loop have to be created and constantly updated to refresh the virtual environment. Thus, the user can see and touch objects in this virtual world with a realistic actions and reactions. The haptic loop must have an update frequency of more than 1000 Hz to make the force felt by the user perceived continuous and uniform. If this frequency is reduced to below 1000 HZ, the haptic handle will start vibrating when hitting the virtual tissue, and the correct force feedback will be disturbed. The graphical loop is linked to the screen refresh rate, which is set to 60 Hz or 60 frames per second.

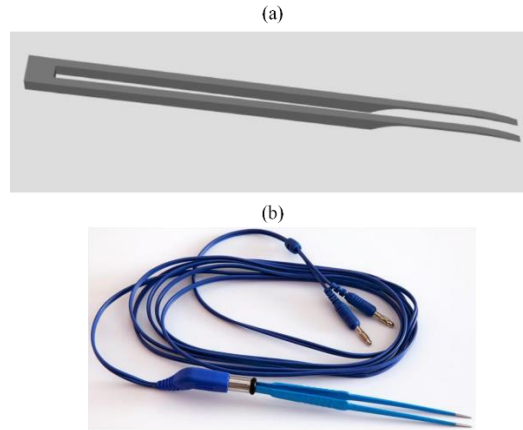

*Figure 1: a) Virtual (modeled) forceps  b) Real forceps based on which the modeling was performed.*

### 2.1.1 Tool

The tools and devices used in the virtual environment must be as similar as possible to the real tools in order for users to connect well with the simulated environment. To do this, we used Solid Works software (Dassault Systèmes SolidWorks Corporation, Waltham, Massachusetts, USA) to design a 3Dmodel of the bipolar coagulation forceps, which is shown in Figure 1-a. The actual bipolar coagulation forceps used in surgery are shown in Figure 1-b. We assigned the bipolar forceps to the haptic handle, so when the handle is moved the modeled instrument is displaced in the same direction and the forces received by the bipolar forceps in the virtual world is transmitted to the handle.

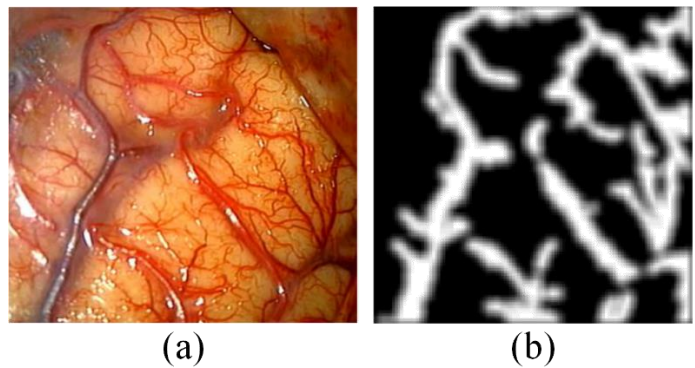

*Figure 2: a) Real image of brain matter, b) extracted veins from the image, and*

### 2.1.2 Brain tissue

Brain tissue was defined as a mesh using a two-dimensional plane consisting of a number of vertices. Figure 2-a shows a sample image used to model the brain tissue. One important tradeoff is between the updating speed and the quality of the tissue model used. If the number of vertices increased, the haptic and graphic loop rate would decrease sharply, and the forces would be in the form of weak taps on the hand. So, the vertices panel size was reduced from 480 by 480 to 120 by 120. The vessels were segmented from the brain image (Figure 2-a) to make the image 3D in order to display them in 3D in the simulated environment.

The image was segmented using a connected components labeling algorithm [17] to extract the veins from the real image. After segmenting the image, thresholding and mean filtering were used to achieve an output like Figure 2-b. This processed image was added to the model to determine the color and the position of the veins. The next step is to simulate how the brain tissue visually and mechanically responds to the touch of the forceps.

### 2.1.3   Visual response

There are two aspects to the visual response of the brain tissue; first, how to show downward deformation in the brain tissue when force is applied to it; second, how the tissue changes color when the current is passed between the tines of the forceps.

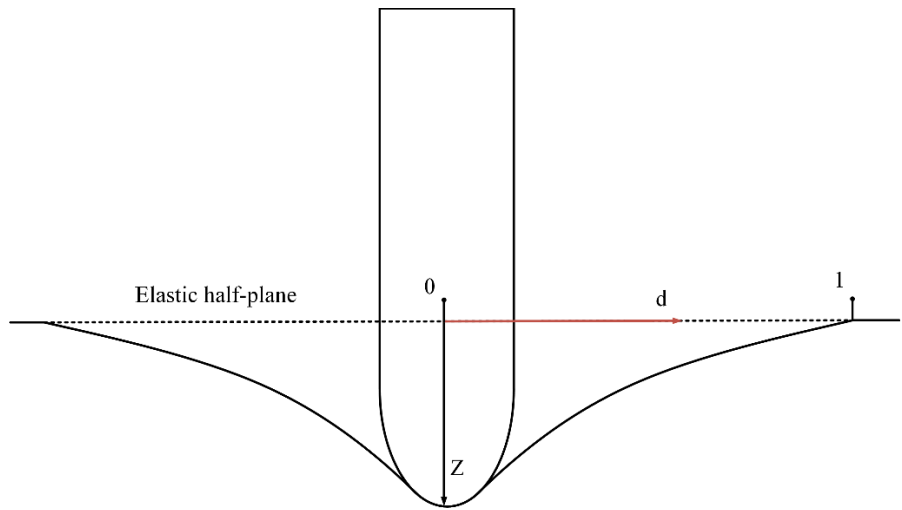

*Figure 3: Indentation model when tissue is under pressure.*

### 2.1.3.1 *Indentation:*

Equation 1 was employed to calculate the indentation of brain tissue receiving pressure from the forceps. We first calculated the distance of the points from the collision point to calculate a relative distance, because the indentation is equal to maximum at the collision point and equal to zero in the indentation radius or farther. Then, the indentation was calculated according to this distance for every point in the indentation radius by Equation 1, where $d_{i,j}$ is the relative distance of the $i, jth$ vertex from the center of collision point. If we want the indentation to be proportional to the location of the forceps on the tissue, we must read its position in the z plane and multiply it by the above expression. It was observed that this dent was not like the natural indentation caused by pressure and exhibited a circular shape. So, the equation was multiplied by an exponentially decaying function. The final equation used in practice was Equation 2, where $z$ is the absolute value of the collision point position in the z plane and $D_{i,j}$ is indentation value of $i, jth$ vertex. Figure 3 shows how this indentation appeared on the brain tissue surface.

$$D_{i,j} = \left(\frac{1}{2} + \frac{1}{2}\cos(\pi * d_{i,j})\right) \tag{1}$$

$$D_{i,j} = z * Exp(-4 * d_{i,j}) * \left(\frac{1}{2} + \frac{1}{2}\cos(\pi * d_{i,j})\right) \tag{2}$$

### 2.1.3.2 *Electrocautery:*

When the electric current passes between the two tines of the forceps, the color of tissue should change accordingly. We modeled this aspect of the simulation according to neurosurgeons' experiences. For this purpose, neurosurgeons were interviewed to elaborate how the colors of the brain tissues change according to the duration of the tissue electrosurgery. If the operation period is too short, the tissue would turn into lighter color, which indicates that the coagulation is unsuccessful. Also, if the operation time is too long, the tissue would turn into very dark color. These features were simulated, and an example is shown in Figure 4.

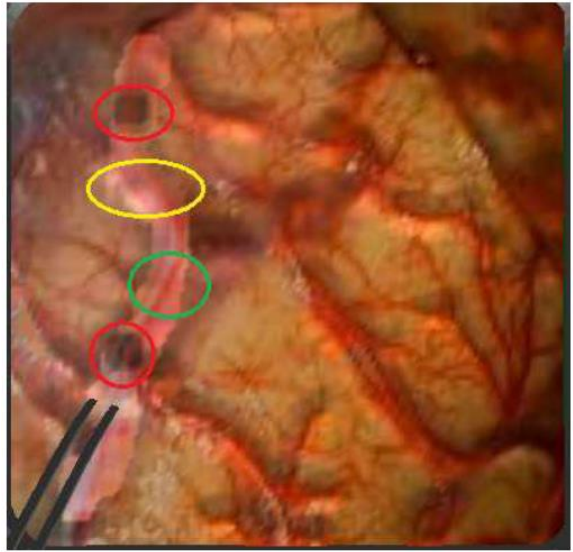

*Figure 4: Different responses of the simulated brain tissue to varying amounts of pressure and duration of electrocautery; red circles are examples of the brain tissue getting damaged, yellow circle is an example of the vessel which is not getting coagulated, and the green circles is an example of proper amount of electrocautery.*

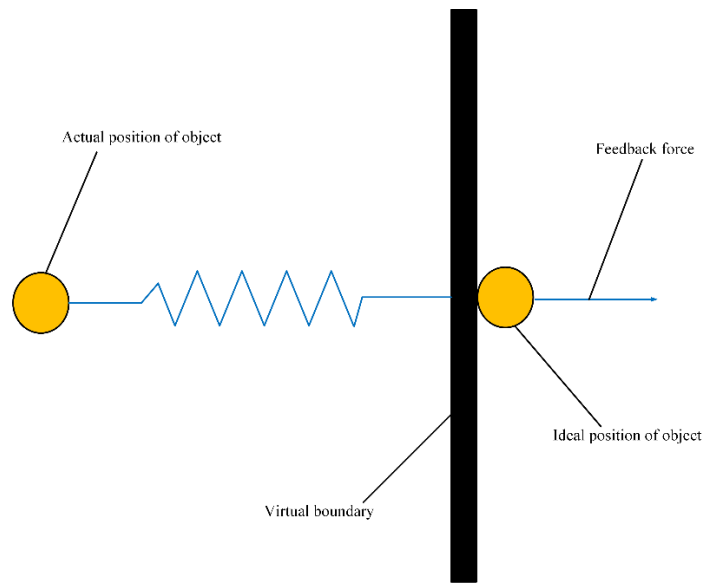

*Figure 5: Spring force model.*

### 2.1.4  Haptic response

The mass-spring-damper model is typically used to model the force of objects. Only the spring model was used to simulate the forces of the brain tissue due to the processing power limits. If a more complex model was used, the updating rate of the haptic loop had to be lowered which would result in suboptimal force feedback. In this model, the location of the object and the texture's boundary are calculated, wherein when the object crosses the virtual texture border, the force is obtained by calculating the distance between the border and the location of the object and multiplying the spring constant of the modeled texture. Figure 5 illustrates the concept of this modeling. The Young's modulus (E) for brain tissue is on average 3 kPa [15]. According to the brain tissue Young's modulus and the developed model in the virtual environment, we can calculate the equivalent spring constant. The brain tissue modeled in the virtual environment is 6 cm wide and 6 cm long. If the brain was modeled by a simple hemisphere, its radius would be approximately 8 cm. Therefore, according to Equation 3, where *L* is the spring's initial length, and *A* is its area, the spring constant (*K*) would be 135 N/m. The friction force was also considered with a friction coefficient of 0.1.

$$K = \frac{A * E}{L} = \frac{0.06 * 0.06 * 3000}{0.08} = 135 \ N/m \quad (3)$$

### 2.1.5 Final adjustments

After implementing the properties mentioned earlier, the color around the brain tissue was changed to bold green to induce a sense of the fabrics used in the operating room. Also, two blue and red buttons were used to begin and end the coagulation process, respectively. These two buttons were placed in a virtual environment to which a magnetic property was associated. The tool sticks to the buttons, and the user can easily activate this feature by pressing the grey-colored button on the haptic handle. After activating the coagulation feature, the same button was considered on the device handle for the coagulation to take place at the desired location. Therefore, after activation, coagulation by the bipolar coagulator takes place anywhere in the tissue by pressing this button.

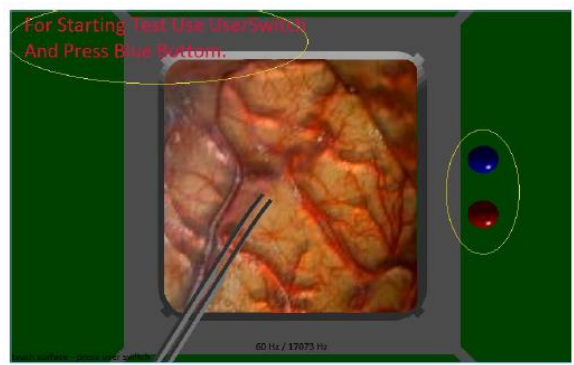

*Figure 6: final simulated environment.*

### 2.1.6 Neurosurgeon's validation

Finally, the simulator was tested by a neurosurgeon at Tehran's Shariati Hospital as a part of the validation process. The surgeon expressed his overall satisfaction with the simulator and the forces that were applied in the form of feedback. Valuable feedbacks were suggested to correct errors and to improve the interaction of the software with user. These feedbacks were taken into consideration prior to finalization of the simulated environment. The final simulated environment is shown in Figure 6.

## 2.2 Experiment scenario

The next step was to design an operation to simulate a surgery task. To this end, a vessel in the simulated brain tissue was selected, and the users were asked to coagulate the target vessel completely with the haptic handle. The coagulation process is repeated ten times for each person, and for each test, the user data is stored to check the learning process during the ten tests. Figure 7 shows the selected vessel from the simulated tissue, which is highlighted in green. The user had to start coagulation from point one to point two and coagulate the entire vein.

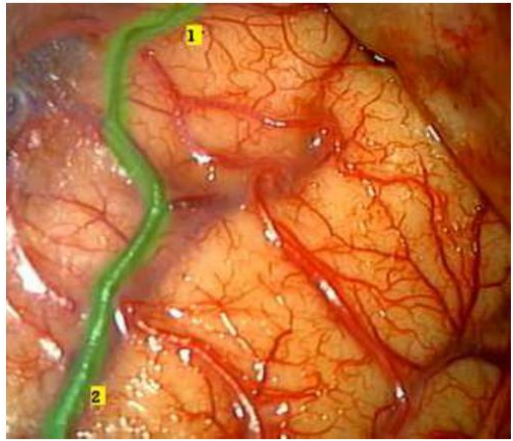

*Figure 7: The path that is asked to coagulate.*

As mentioned earlier, the amount of coagulation depends on the force applied by the user and the time the user holds the forceps. If the user performs too much coagulation, the texture turns into black, indicating excessive clotting and brain tissue damage. On the other hand, if the user coagulates less than necessary, the surface becomes brighter than the original intensity. Thus, the user would receive visual feedback to adjust the coagulation rate. This process was explained to candidates as the user guidance to employ this tool.

The test is repeated ten times for each candidate. Moreover, for better training of the candidates, an additional trial was considered for training the candidate, which was done with the guidance of the test supervisor. Thereafter, the supervisor did not intervene during the other ten experiments. The test procedure was as follows: The candidate first pushes the blue button with the handle to start the test. After starting the test, the user must coagulate the vessel and afterward the red button should be activated to complete the test.

During the test for each candidate, time, forceps position, coagulation rate, and force are stored from the time the candidate pushes the blue button at the beginning of the test until the test is finished by pressing the red button. This data was stored separately for each experiment and person to be examined in the next step.

## 2.3 Evaluation

Different evaluation metrics were defined to measure the performance improvement of the candidates as follows

1- Length of the tool tip path (route): each candidate covers a path from the starting point to the endpoint to perform the test. The size of this path was calculated as a performance metric.
2- Duration of test: time of the operation performed by the volunteer. This time was calculated from the start of the trial until the end.
3- Average force applied to the tissue: in each experiment, the average force applied by the user to the simulated brain tissue was calculated.
4- Standard deviation of the applied force: after calculating the average force applied to the tissue, its standard deviation was also determined.
5- Percentage of applied force error: this criterion was defined as the force applied by the volunteer to the specified vessel tissue divided by the total forces applied to the surrounding tissue. In the experiment, the candidates had to coagulate only the vein, and if they mistakenly coagulated the brain tissue, it would be considered as an error.
6- Percentage of brain tissue coagulation error: this metric indicates the coagulation rate of non-target tissue to the total coagulated tissue.
7- Percentage of the vein not coagulated: the area not coagulated in the vein to the area of the entire vein.

In addition, the candidates were divided into two groups of gamers and non-gamers. The results achieved by these two groups were compared to understand the effect of playing computer games on surgical skills.

## 3 Results and Discussion

In this section, the experiment results of learning process for twenty students are presented. The experiment was performed with full consent of the candidates. The consent form described the project's purpose, the benefits of participating in the test, and the confidentiality of personal information.

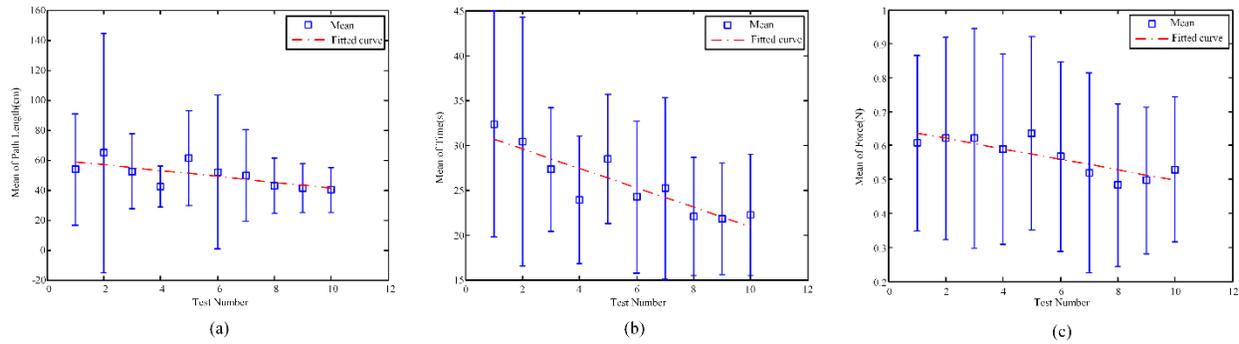

*Figure 8: These graphs illustrate the mean and standard deviation of the a) path length, b) test time, and c) force applied by the volunteers*

### 3.1 length of the path

One of the criteria defined to examine the learning process is the length of the path covered within the coagulation process. Although this is not a good criterion for checking the accuracy of the task assigned to the volunteers, we expected that the path length would decrease when the volunteers repeated the test. The results shown in Figure 8-a illustrates the mean and standard deviation of the path length for the subjects. It is clearly seen that the mean and standard deviation in the first experiments is more than those of the last experiments. The fitted line demonstrates the constant reduction in the path length treated by the candidates. A t-test was used to investigate the significance of the differences between the first and last experiments, which demonstrated statistically valid improvement with p-value of 0.0063. This p-value showed a substantial improvement between the first and last experiments for the entire candidates.

### 3.2 Duration of test

The time of test metric, similar to the path length metric, does not accurately reflect the performance of volunteers. However, it is expected that the time for the final tests will decrease as the volunteers gain experience in conducting the test. Figure 1-b shows the mean and standard deviation for each experiment, wherein the mean and standard deviation are generally decreasing. It should be noted that the red line has been fitted on the average time. The p-value between the first and last experiments was 0.0022, which indicates statistically significant difference between first and last experiment. Overall, the total time spent on each experiment decreases, and at the last three experiments the experiment time converged to a constant number.

### 3.3 Average force applied to the tissue

Brain tissue is one of the most sensitive, so it should be treated carefully and with low force within surgery. Volunteers are expected to deal with tissue with little force or standard deviation. According to the Figure 8-c, the applied force was reduced as the number of tests proceeded, though the differences between early and latest tests is not remarkably high. The mean applied force was about 0.6 N. The P-value calculated between the first and last experiments for this metric was 0.0338, which indicates a significant difference between the two experiments.

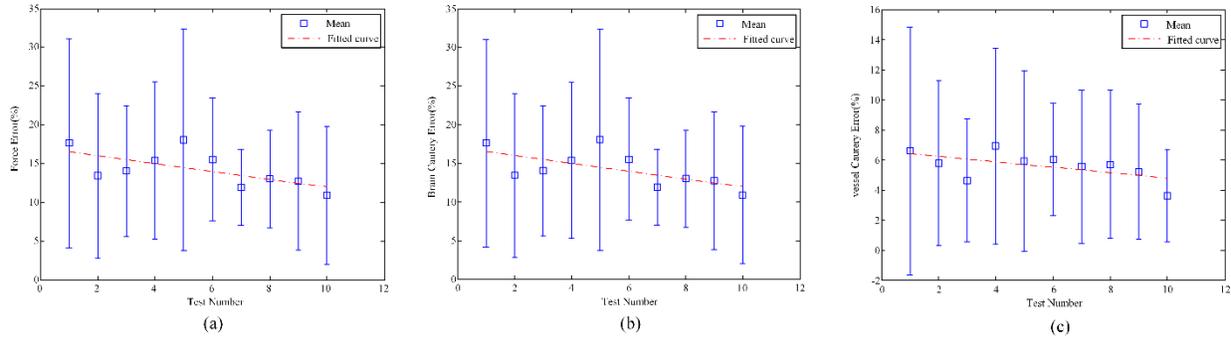

*Figure 9: These graphs illustrate the mean and standard deviation of the percentage of applied force, brain tissue coagulation, and not coagulated vein errors generated by the volunteers*

### 3.4 Percentage of applied force error

It was expected that the volunteers committed less errors within the tests and only apply force/treatment to the designated vein. The graphs in Figure 9-a shows the mean and standard deviation of the force error in percentage. The fitted line shows a negative slope, which indicates a constant reduction in force error. Moreover, the standard deviation of the error is decreasing. The p-value calculated between the first and last experiments was equal to 0.0350, which indicates a significant reduction in force error.

### 3.5 Percentage of brain tissue coagulation error

One of the aims of training was that the volunteers spare the brain tissue from the coagulation. The graphs in Figure 9-b display the mean and standard deviation of the brain tissue coagulation errors in percentage ratio of the brain tissue coagulated to all the tissue coagulated. The mean and variance of this error exhibited overall a decreasing trend. The p-value obtained between the first and tenth experiments was equal to 0.0274, which confirmed a significant difference between the first and tenth experiments.

### 3.6 Percentage of the vein not coagulated

The aim of training is to coagulate the entire vein. Thus, the percentage of coagulated to non-coagulated vein would reflect the efficiency of the training. This error should eventually be minimized as the volunteers' training proceeds. The graphs in Figure 9-c present the mean and standard deviation of the percentage of non-coagulated vein. The fitted line on the mean values exhibited a negative slope, and the overall standard deviation decreased as the training proceeded. The p-value between the first and last experiments was 0.0081.

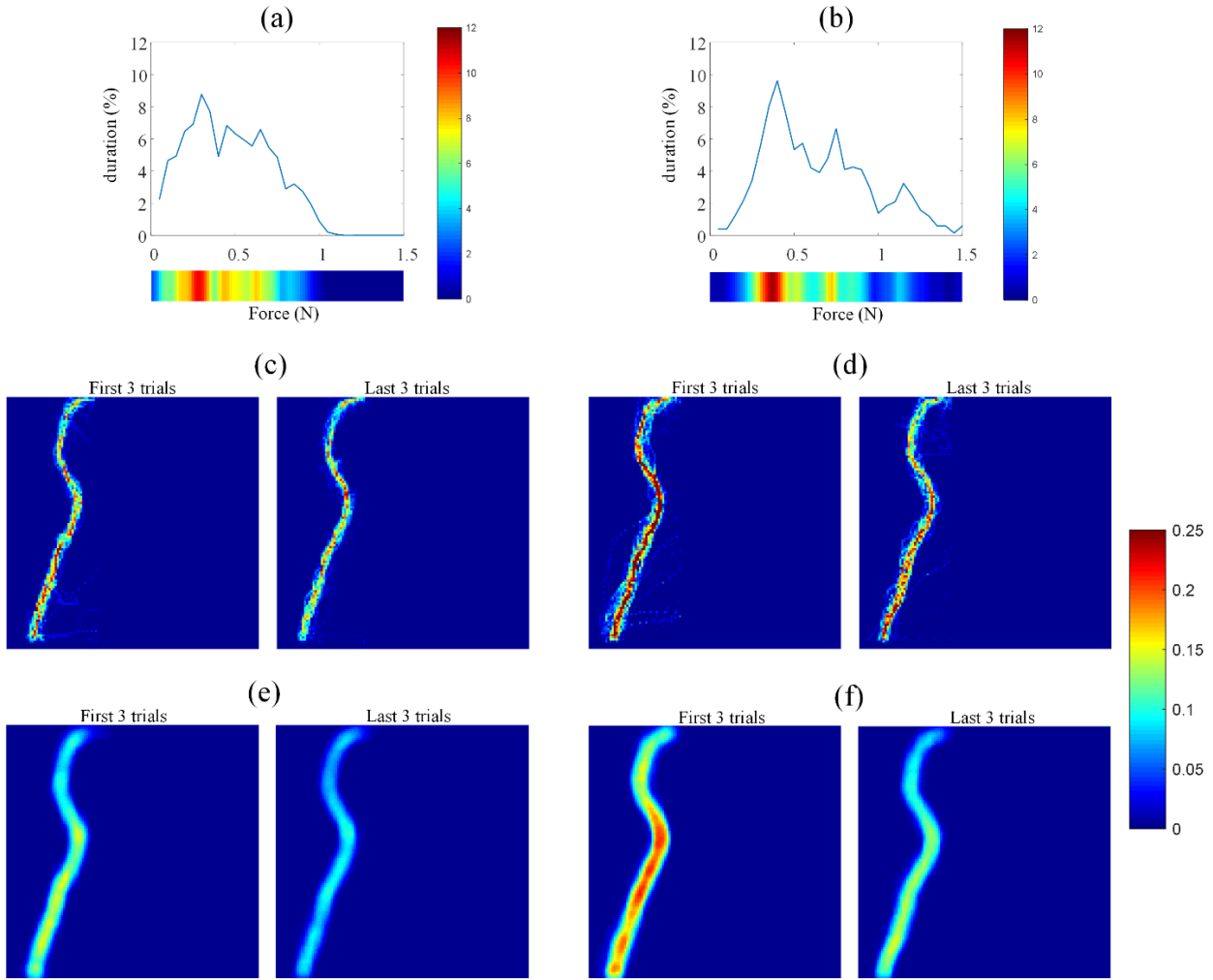

*Figure 10: Average force intensity in a) the gamer group and b) the non-gamer group. Heat map of c) the mean force applied to vertices of brain tissue by the gamer group, d) the mean force applied to vertices of brain tissue by the non-gamer group, e) mean coagulation rate for every vertex in the gamer group, and f) mean coagulation rate for every vertex in the non-gamer groups*

## 3.7 Comparative study

In the questionnaire filled by the candidates, a question was asked whether they played computer games. The 20 volunteers in the experiment were divided into two groups: gamers (10 persons) and non-gamers (10 persons). We ran a comparative study between these two groups.

In electrocautery, avoiding excessive force to brain tissue is very crucial to impose minimum possible harm to the healthy tissues. The amount of force applied by these two groups are shown in Figure 10-(a,b). This force intensity diagram shows the percentage of applied force to the target in respect to the entire force. These graphs show that the gamers group did not put pressures above 1 N and had an overall less average force.

The heat map of force and electrocautery path averaged over the entire candidates in the gamer and non-gamer groups are presented in Figure 10-(c,d,e,f) (the average of 3 first and 3 last trials are provided separately). The mean force was reduced as the trials proceeded in both groups. It is also evident that the gamer group applied less but more uniform force. The same trend was observed in coagulation path, wherein the gamers had better performance after the training.

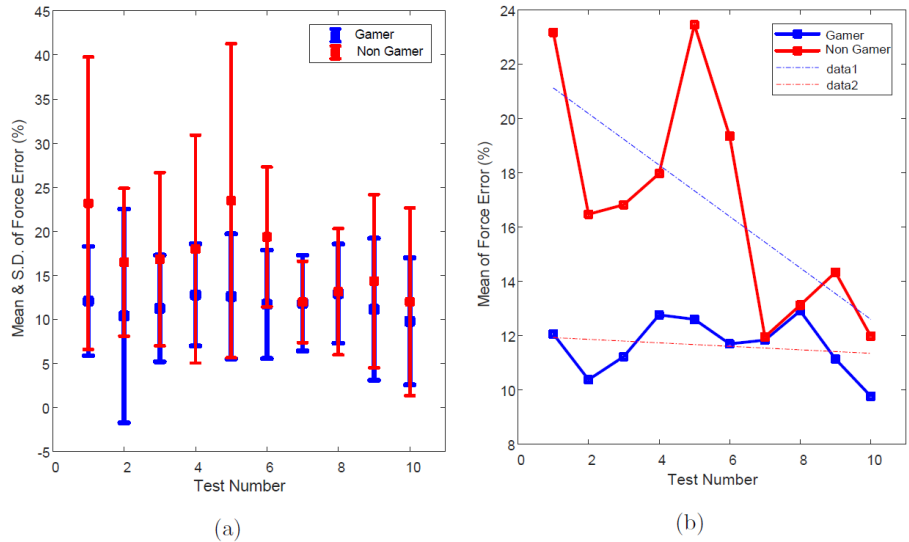

*Figure 11: a) The mean and standard deviation of the erroneous force applied by the gamer and non-gamer groups.*

The mean and standard deviation of the erroneous force applied to the non-vascular brain tissue instead of the vessel is depicted for the two groups in Figure 11. The gamer group had a relatively lower mean and standard deviation than the non-gamer group. The gamer group started the experiment with a lower mean and standard deviation than the non-gamer but completed the tests with similar results as their first test. However, the non-gamer group started with a higher error rate and standard deviation. However, in the final experiments they achieved the results of the gamer group. So, the trend of error reduction for the non-gamer group (considering the slope of fitted line) was faster than the gamer group. The p-value between these two groups was 0.0001, which demonstrated that these two groups have a significant difference. These results indicated that playing computer games would increase the hand control, focus, and reflex.

## 4. Conclusion

In this project, a neurosurgery simulator was designed to simulate brain tissue deformation/response to the force applied within the operation. This simulator features coagulation of blood veins in the brain. This simulator was used to evaluate people's learning process within the brain surgery. Our observation demonstrated that candidates performed better as the trials proceeded with significantly less errors. Moreover, the gamer candidates exhibited higher accuracy and less errors within the treatment compared to the non-gamer group. The simulator could play a significant role in the training of the surgeons and accelerates the process of training.

## Acknowledgments
We thank all the volunteer students who participated in these studies. We would like to particularly thank Alireza Khoshnevisan from Shariati Hospital in Tehran, whom we consulted in designing the simulator and who was kind enough to help us evaluate the simulator.

## Supplementary material
A video demonstration of the simulator in use can be found at: https://youtu.be/OEBFIWKelBI